\documentclass[preprint]{aastex}
\newcommand{\be}{\begin{enumerate}}
\newcommand{\ee}{\end{enumerate}}
\newcommand{\bi}{\begin{itemize}}
\newcommand{\ei}{\end{itemize}}
\newcommand{\bd}{\begin{description}}
\newcommand{\ed}{\end{description}}

\newcommand{\myarcmin}{^\prime\mskip-5mu}

\newcommand{\rosat}{{\it ROSAT}}

\newcommand{\asca}{{\it ASCA}}

\slugcomment{Accepted for publication in {\it The Astrophysical Journal}}

\begin{document}
\title{A New ASCA and ROSAT Study of the Supernova Remnant: G272.2$-$3.2}

\author{Ilana M.~Harrus\altaffilmark{1}}
\affil{NASA/USRA Goddard Space Flight Center, Greenbelt MD 20771}

\author{P. O. Slane and R. Smith}
\affil{Harvard-Smithsonian Center for Astrophysics, Cambridge MA 02138}

\and
\author{J. P. Hughes}
\affil{Rutgers University, Piscataway NJ 08854}

\altaffiltext{1}{imh@milkyway.gsfc.nasa.gov}
 
\begin{abstract}

\par
\noindent
G272.2$-$3.2 is a supernova remnant (SNR) characterized by an apparent 
centrally brightened
X-ray morphology and thermally dominated X-ray emission. 
Because of this combination of Sedov-type (thermal emission) and non-Sedov
type (non-shell like morphology) features, the remnant is classified as 
a ``thermal composite'' SNR. 
This class of remnant is still poorly understood due in part to the
difficulties in modeling accurately all the physical conditions which
shape the emission morphology. \\
In this paper we present a combined analysis of data from the \asca\
and {\it ROSAT} satellites coupled with previous results at other wavelengths.
We find that the X-ray emission from 
G272.2$-$3.2 is best described by a non-equilibrium ionization 
(NEI) model with a temperature around 0.70~keV, an ionization timescale 
of 3200~cm$^{\rm -3}$ yr and a relatively high column density
($N_{\rm H}\sim 10^{\rm 22}$ atoms cm$^{\rm -2}$). 
We look into the possible explanations for the apparent morphology of 
G272.2$-$3.2
using several models (among which both 
cloud evaporation and thermal conduction models). 
For each of the models considered we examine all the implications on the 
evolution of G272.2$-$3.2. 
\end{abstract}


\section{ Introduction }
\par
\noindent
Supernova remnants are potentially 
powerful tracers of the complete history of their 
progenitor star. Their study may provide information not only on the 
amount of material ejected by the star
during the course of its life, but also on the composition of the 
interstellar medium (ISM) in which the supernova-explosion blast wave 
propagates.\\
In a simplified model of supernova evolution, this blast wave 
propagates through a homogeneous ISM. 
When the amount of ISM material swept-up by the wave becomes
comparable to the mass contribution due to the ejecta, the remnant
enters the adiabatic, also called Sedov-Taylor, 
phase of its evolution \citep{tay50,sed59}.
During this phase, the radiative cooling is negligible and the overall 
energy of the SNR remains roughly constant.
For an SNR evolving according to this model, and in
 absence of any compact object created during the supernova explosion, 
the expected X-ray profile is a shell, and its 
spectrum is characteristic of collisional excitations in a 
plasma with a temperature around 1~keV.  
Indeed, a large number of SNRs 
present this kind of morphology (for example the 
Cygnus Loop). \\
However, most of the SNRs do not conform to this simple 
picture and they present characteristics which substantially differ from this 
ideal model. 
One of the many subgroups 
consists of SNRs which appear centrally bright in X-rays and 
present spectra dominated by thermal emission. 
It is widely believed that this peculiar morphology is linked to 
inhomogeneities in the ISM (one model we use later favors small cold
clouds) but few models can account for the measured temperature
profiles across remnants in this class. 
We present here an analysis of recent X-ray observations of G272.2$-$3.2,
including an attempt to reproduce both the morphology and the
temperature profile of the remnant. 
We first examine the existing data on G272.2$-$3.2, in particular
radio and optical observations, and summarize the results obtained from 
these data. 
\noindent 
We then describe briefly the X-ray observations used in our analysis. 
Our image and spectral analysis are presented in \S~3, along with a
study of the temperature profile and the theoretical modeling of
the X-ray morphology. 
\S~4 examines the consequences of the results presented in the previous
paragraph. There we discuss and summarize  the main points of the paper.

\section{From Radio to X-Ray: Pre-\asca\ knowledge on G272.2$-$3.2}
\noindent
G272.2$-$3.2 was discovered in the \rosat\ All-Sky Survey \citep{gre93}. 
It presents a centrally filled X-ray morphology and a thermally dominated 
X-ray spectrum. 
The temperature, as determined 
by the \rosat\ PSPC observation is relatively 
high (between 1.0 and 1.5~keV for a 1$\sigma$ confidence level). No
significant variation of the temperature could be measured with the X-ray 
data available from the \rosat\ PSPC \citep{gre94}.\\
Optical observations \citep{win93}, made soon after the 
announcement of the discovery, confirm the nature of the nebulosity as 
a supernova remnant and the shock-heated
nature of the emission. 
In particular, both the measured [S{\sc ii}]/H$\alpha$ ratio 
and the detected emission from [N{\sc ii}]~658.3-nm 
and [O{\sc ii}]~732.5-nm are typical of SNRs. 
One of the fainter filaments is located at 
09$^{\rm h}$06$^{\rm m}$40$^{\rm s}\!.$4, 
-52$^\circ$06$^\prime$42$^{\prime\prime}$ (J2000), about 
1$^{\prime}$ north-west of the X-ray emission center. 
Brighter filaments with a similar [S{\sc ii}]/H$\alpha$ are detected at 
09$^{\rm h}$06$^{\rm m}$12$^{\rm s}\!.$4, 
-52$^\circ$06$^\prime$52$^{\prime\prime}$. There is no evidence for diffuse 
emission in the continuum images of the field. \\
Detailed radio observations \citep{dun97} have introduced a 
wealth of new data about this SNR. 
The radio observations were conducted at multiple frequencies and baselines 
using the Parkes radio telescope, the Australia Telescope Compact Array
(ATCA) and the Molonglo Observatory Synthesis Telescope ({\it MOST}).
G272.2$-$3.2 is of such low surface brightness ($\sim$ 1~mJy at 
843~MHz)
that even the sub-arcminute {\it MOST} observations could only map the 
brightest parts of the remnant. G272.2$-$3.2 presents an 
incoherent filamentary structure with diffuse, non-thermal emission of very
low surface brightness as well as bright ``blobs'' which correlate 
well with the brightest optical filaments. This is a good indication
that optical and radio emission emanate from the same regions. These regions
are likely to be the ones where the shock  
interaction with the ISM is the strongest. 
\setcounter{footnote}{0}
We show in Fig.~1 a {\it MOST} radio image extracted from \citet{dun97} 
on which the authors show the position of the optical filaments. 
There is no evidence for a clean
``shell-like'' morphology although the remnant is roughly circular with a 
20$^\prime$ diameter and has a steep non-thermal radio spectral index
(0.55$\pm$0.15), which is more typical of shell-like remnants than
plerions. This  implies that the diffuse emission detected within 
G272.2$-$3.2 is
 most probably due to shock-accelerated electrons.
The spectral index varies relatively little across the remnant. 
There is no evidence of polarized emission from the remnant and \citet{dun97} 
argue that this depolarization may be the result of
turbulence occurring on angular scales of the order of~1$^{\prime}$. 
There is no evidence of  pulsar-driven nebula.\\
\noindent
There are data from the Infrared Astronomical Satellite ({\it IRAS}) survey
at 12, 25, 60 and 100~$\mu$m. 
Infrared observations are unique in that they provide direct
information on the dust present in the ISM and so they 
can serve as a powerful tool for temperature diagnostics. 
We have extracted the {\it IRAS} data\footnote{Data extracted from the
HEASARC-SkyView site at: http://skyview.gsfc.nasa.gov/} 
at all four wavelengths and
present our findings in the next section.\\
There are CO maps of the G272.2$-$3.2 sky region at a resolution of
18$^{\prime}$ (larger than the remnant). In this 
direction in the sky, velocity changes very little along the line of
sight adding to the  difficulty of constraining the distance to the
remnant. The only significant CO emission is measured at velocities 
between -20~km~s$^{\rm-1}$ and 20~km~s$^{\rm-1}$
(T. Dame - private communication). \\
\noindent
At a distance greater than 2~kpc, the large galactic latitude of 
G272.2$-$3.2 implies
a distance below the plane larger than 110~pc. Using the distance 
distribution of SNRs in the Galaxy \citep{mih81} and 
their galactic latitude \citep{gre98}, we estimate that less than 10\% of 
galactic SNRs are located at a greater distance from the plane. 
This results is in agreement with 100~pc upper limit given by \citet{all85} 
for 
the distance distribution of stars above the plane of the Galaxy and suggests
a distance of about 2~kpc to  G272.2$-$3.2. \\
\noindent
This estimate, although solely based on a statistical analysis, turns out to 
be in good agreement with a value of 
$1.8^{\rm +1.4}_{\rm -0.8}$~kpc based on the measured X-ray column density of 
$N_{\rm H} = 4.6\pm3\times 10^{\rm 21}$atoms~cm$^{\rm -2}$ \citep{gre94}.
\noindent
We also derive an estimate of an upper limit for the distance 
to the remnant. 
Using stars within a distance of 2~kpc, \citet{luc78} finds an optical color 
excess with distance of roughly 0.2 mag~kpc$^{\rm -1}$ in the direction of 
G272.2$-$3.2.
Using the relation between color excess and column density 
($N_{\rm H}=5.9\times 10^{\rm 21}\times E_{\rm {B-V}}$~atoms~cm$^{\rm -2}$)
of \citet{pre95}, one gets an upper limit of about 10~kpc for 
the distance to G272.2$-$3.2. \\
\noindent
In view of all the uncertainties on the distance measurement, 
we adopt an ``intermediate'' distance scale of 5~kpc 
in the following computation and will 
examine the consequences of this estimate on the 
different dynamical states of the remnant. One must keep in mind 
that this value is mainly used as a scaling factor; we will keep the 
distance variation explicit in all our 
computations to allow easy computations at other distances 
of the physical quantities derived.

\section{\asca\ observations}
\subsection{Spatial Analysis}
\subsubsection{Images}
\vspace{-0.1cm}
\noindent
\asca\ carried out one observation of G272.2$-$3.2 on 
1995 January 20 at a nominal pointing direction of 
09$^{\rm h}$06$^{\rm m}$43$^{\rm s}\!.$2, 
-52$^\circ$06$^\prime$14$^{\prime\prime}\!.$4 (J2000).
After applying the standard cuts on the data, we
generated exposure-corrected, background-subtracted merged images of the 
GIS and SIS data in selected spectral bands. Background was determined 
from the weighted average of several nominally blank fields from high-galactic 
latitude observations with data selection criteria matched to those used for
the SNR data. Exposure maps were generated from the off-axis effective-area
calibrations, weighted by the appropriate observation time. 
Events from regions of the merged exposure map with less than 10\%  
of the maximum exposure were ignored. Merged images of the source data,
background, and exposure were smoothed with a Gaussian of 
$\sigma$= 30$^{\prime\prime}$ for both the low-energy band (0.5--4.0 keV) and 
the high-energy band (4.0--10.0 keV).
We subtracted smoothed background maps from the data maps and divided by the
corresponding exposure map. Fig.~2 shows the results obtained for 
the two detectors in the low energy band. The SNR is not detected 
above 4.0~keV and confirms the result from radio data that there is no sign 
of a pulsar-driven nebula. 
To examine the morphology  of the remnant in more detail, we have generated 
SIS images in narrow energy bands in an attempt to isolate contributions  from 
separate elements. We have minimized the continuum component by
subtracting the average contribution from a 
small range of energy around each line imaged. 
In Fig.~3 we show SIS images of G272.2$-$3.2 
in the three narrow energy bands [1.20 keV--1.35 keV],  
[1.70 keV--1.90 keV] and [2.25 keV--2.50 keV]
corresponding to Mg XI, Si XIII, S XV lines respectively. 
The minimum visible in the center of all the images 
is due to an instrumental effect and corresponds to 
the location of the wide gap between CCDs in the SIS detector.
There is a strong correlation between the Mg XI energy band map and 
the broad energy image. In the next section (spectral analysis) we will
examine how this effect translates to 
a higher value of magnesium abundance inside 
the region of maximum emission. \\
\subsubsection{Comparison with existing data}
\noindent
To get yet a better estimate of the morphology of this remnant, we have 
analyzed \rosat\ data  from both the High Resolution Imager (HRI) and the
Position Sensitive Proportional Counter (PSPC).
Both sets of data 
were cleaned according to the standard 
prescription to study extended sources \citep{sno94}. 
\setcounter{footnote}{0}
The software\footnote{Available via anonymous ftp at ``legacy.gsfc.nasa.gov''.}
computes the contributions from the different backgrounds (solar scattered
X-rays, high-energy particles, long and short term enhancements) and subtracts
them from the data. Similar corrections are made for the HRI although
contamination from those backgrounds is known with less accuracy.
Fig.~4 shows the result of this procedure for the \rosat\
HRI (the result is similar for the PSPC). 
A bright spot at the western edge of G272.2$-$3.2
is seen with a flattening of the shell 
at this location which coincides
with the bright optical filament detected by \citet{win93}.
One explanation for this bright spot is that the expanding shock is 
encountering a density gradient in the local ISM. It could also well
be that the shock has engulfed a cloud, slowed down, and that the cloud
is being evaporated. At a distance of 5~kpc, the angular size of about
2$^\prime$25$^{\prime\prime}$ implies a cloud 3.5~pc in size. Using the 
results from the spectral analysis (see next section)  we deduce 
a density between 0.42 and 0.70 cm$^{\rm -3}$ for the clouds.      
The soft emission of that region is compatible
with both explanations. \\
\noindent
One useful indicator of the presence of dust in the galaxy is in the 
infrared energy band. Using data from the {\it IRAS} survey, 
\citet{sak92} studied the infrared
emission from 161 galactic 
SNRs. 
They argue that young remnants tend to have strongest  fluxes at 12$\mu$m 
and 25$\mu$m while somewhat older remnants have their strongest 
emission at longer wavelengths. 
G272.2$-$3.2 was not part of this study although its  
{\it IRAS} data
show strong emission at 
60 and 100$\mu$m (15 and 70 MJy~str$^{\rm -1}$ respectively) 
probably associated with the shock heated dust in the ISM. 
There is no significant emission at the two smallest wavelengths, a possible
indication of a low dust temperature. 
One of the techniques commonly used 
to distinguish legitimate SNR emission 
from H{\sc ii} regions and eliminate potential calibration or 
normalization problems, consists of studying the ratio of 
maps (60$\mu$m/100$\mu$m; 12$\mu$m/25$\mu$m) as
an indicator of the respective contributions. 
This technique, applied 
with success for the Cygnus Loop \citep{sak92}
has the advantage of being both simple and 
free from a lot of theoretical assumptions
(about grain emissivities for example) that would otherwise
complicate the interpretation.\\
\noindent
We computed the ratio $F_{\rm 60}$/$F_{\rm 100}$ from the remnant 
(using the X-ray image to define its extent).
We find that $F_{\rm 60}$/$F_{\rm 100}$~$\sim$~0.2 which is comparable
with what was measured for Vela XYZ \citep{sak92} 
but still lower than almost
all the other ratios quoted (the highest being 2.45 for Kepler's
SNR). This low value of the  $F_{\rm 60}$/$F_{\rm 100}$ emission is consistent 
with the low dust temperature hinted at by the lack of emission at 12$\mu$m 
and 25$\mu$m.  
Fig.~5 shows the smoothed (with a Gaussian of 3$^\prime$ ) 
image of the 60$\mu$m/100$\mu$m ratio and contours from a close-up of this 
image shown with the \rosat\ HRI image superposed.
The correlation with the brightest part of the SNR (the western
part of the remnant) is obvious and in complete agreement with the optical,
radio and X-ray data.  It is remarkable not only that the SNR is
so obviously detected in the {\it IRAS} data (and its very high
background level), but that the agreement with the HRI image is so
good. 
In the next section, 
we analyze spectra extracted from the brightest region and the rest of the
remnant. 
The results from both spectral and spatial analysis are then used to 
form the global picture of the remnant. 

\subsection{Spectral Analysis}
\vspace{-0.1cm}
\noindent
Depending on the age of G272.2$-$3.2 and on the pre-shock medium density, 
non-equilibrium ionization effects can become important \citep{ito79}. 
In this case, a simple equilibrium collisional plasma 
emission model (Raymond \& Smith 1977 -- CEI model) can no longer 
be applied
to reproduce the expected X-ray spectra. One has to take into account the 
fact that the ions are not instantaneously
ionized to their equilibrium configuration at the temperature of the
shock front.
This model has, in addition to the temperature, an additional parameter 
which describes the state of the plasma ionization.
This ionization state 
depends on the product of electron density and age and we define the 
ionization timescale as  $\tau_i \equiv n_et$. 
We have used an NEI model \citep{hug85} 
keeping all the elemental abundances 
at their value given in \citet{and89} except when explicitly
mentioned. 
In order to check the possibility of weak non-equilibrium ionization effects
(for the dynamically older parts of the remnant), 
we have run both models (CEI and NEI) and compared the results from our 
spectral analysis.

\subsubsection{Results from the Spectral Analysis}
\noindent
We have carried out several spectral analyses using the different data sets
available and then combined them in order to obtain a 
general picture of the remnant. 
We have already seen that the remnant is
undetected above 4~keV; all channels above that energy are ignored in the
following analysis. 
After a first fit using only the \asca\ data, we added data from the 
\rosat\ PSPC to constrain $N_{\rm H}$,  
the value of absorption along the line of
sight, using the cross-sections and
 abundances from Morrison \& McCammon (1983). 
\noindent 
In all the analyses, the data were extracted from the \rosat\ PSPC
observation and then spatially matched to that of \asca.
\setcounter{footnote}{0}
We added a gain shift to both 
GIS~2 and GIS~3 (the same value for both detectors)
according to the prescriptions from the 
calibration data analysis done by the \asca-GIS team\footnote{see http://heasarc.gsfc.nasa.gov/docs/frames/asca$_{-}$proc.html 
for more informations on calibration issues.}.
An initial \asca\ analysis was done on the complete remnant. We extracted
a total of 25000 GIS events from a circular  region encompassing 
the total emission region. The SIS was 
not used for the full SNR spectrum  because the remnant covers more than one 
chip. 
We found that although it was impossible to describe accurately 
the complete SNR using a single model (CEI or NEI), 
the quality of the fit improves dramatically (from a 
 $\chi^2$ of 870 to 514, for 325 degrees of freedom) in going from a CEI to an
NEI model. 
We found a common gain shift (for both GIS~2 and 3) of -3.3\% , 
consistent with the results found by the GIS calibration team. \\
The spectrum shown in Fig.~6 shows strong residuals at the silicon and
sulfur energy lines. Neither the NEI nor the CEI can accurately model 
these two strong emission features.\\
In the next step of the analysis, we included  data
from the \rosat\ PSPC  data in the fit. 
In this case the column density is 
$N_{\rm H} = 1.12\pm0.02\times 10^{\rm 22}$ atoms~cm$^{\rm -2}$, the 
ionization timescale is $2150^{\rm +320}_{\rm -360}$~cm$^{-3}$~years,
 while the temperature is  $kT = 0.73^{\rm +0.03}_{\rm -0.04}$ keV (associated
with a $\chi^2$/$\nu$ of 691.95/354).
Both ionization timescale and the temperature of the plasma are compatible with
the results found in the previous analysis but the 
column density is  smaller.
The total unabsorbed flux between 0.2 and 3.0 keV is 
$2.35\pm0.15\times10^{\rm -10}$~ergs~cm$^{-2}$~s$^{-1}$. The
fit for the complete remnant is shown in Fig.~6 and all the results are 
given in Table.~1. We note that the errors quoted for the measured parameters 
are underestimates in that they correspond to a fit with a large 
$\chi^2$/$\nu$. \\
\noindent
In order to get a more quantitative picture of the remnant, we have separated
it into ``A'' and ``B'', two non-overlapping regions of emission. 
The ``A'' region 
is chosen to encompass the bright western part of the remnant (see Fig.~2 
--left panel-- or the following
paragraph for a definition of the region) while 
the ``B'' spot is taken from the other region of the SNR, where no optical 
filaments have been observed. 
\subsubsection{Study of the ``A'' Region}
\noindent
We have extracted events from the ``A'' region defined in the 
\rosat\ PSPC as a $\sim$ 2 $\myarcmin$ radius 
circle centered at 09$^{\rm h}$06$^{\rm m}$11$^{\rm s}\!.$4, 
-52$^\circ$05$^\prime$34$^{\prime\prime}\!.$4 (J2000).\\
The same region is then selected for the \asca\ SIS and GIS.
Unfortunately the region is at the edge of
the \asca\ SIS and the circular region of extraction is
truncated to take this into account.  After background subtraction, 
the count rate is $0.084\pm0.002$~cts~s$^{\rm -1}$,
$0.024\pm0.001$~cts~s$^{\rm -1}$, and 
$0.017\pm0.001$~cts~s$^{\rm -1}$ in the 
\rosat\ PSPC and \asca\ SIS/GIS respectively (we have averaged the values
for SIS 0 and SIS 1 as well as those for GIS 2 and GIS 3).\\
We model the five spectra using the NEI model mentioned above. 
As previously a gain shift is included in the  analysis of 
GIS~2 and GIS~3 data.\\ 
The resulting $\chi^2$/$\nu$
is 248.35/133.  All the results for region ``A'' are consistent with the ones
found for the complete remnant when fit with a NEI thermal model. 
We find a column density of 
$N_{\rm H} =  
1.17^{\rm +0.02}_{\rm -0.06}\times 10^{\rm 22}$~atoms~cm$^{\rm -2}$, 
an ionization timescale of $760^{\rm +830}_{\rm -70}$~cm$^{-3}$~years 
and a temperature of $kT = 0. 86^{\rm +0.08}_{\rm -0.22}$~keV. 
We found a gain shift of -3\% consistent with the results found 
previously (see results Table.~2a.). 
In comparison a fit using a CEI model \citep{mew85,mew86,kaa92}, 
leads to a worse fit ($\chi^2_{\rm r}$ = 1.96).
As we defined region ``A'' to be coincidental with the region of enhanced
Mg~X lines (see circle in Fig.~3), we added 
magnesium abundance as an extra parameter in the fit 
and check for any statistically significant drop in $\chi^2$. 
With this one extra parameter, the $\chi^2$/$\nu$ is now 
176.07/129. The probability to exceed this value per chance is 0.003 compared 
to 5$\times 10^{\rm -9}$ for the previous fit. 
The column density is consistent with the previous result 
($N_{\rm H}$ = 
$9.5^{\rm +0.7}_{\rm -1.0}\times 10^{\rm 21}$~atoms~cm$^{\rm -2}$), 
and so are the temperature 
($kT = 1.00^{\rm +0.35}_{\rm -0.08}$ keV)  and the ionization timescale 
($1385^{\rm +1860}_{\rm -550}$~cm$^{\rm -3}$ years).
The SIS detectors (the more sensitive instruments 
for measuring any abundance variation)  do show a departure from the cosmic
value (the smallest acceptable values range between 1.3 to 2 times 
solar abundance) but one has to keep in mind that the data were taken in 
a 4-CCD mode with degraded resolution and non-optimized calibration. 
All the other detectors yield spectra indistinguishable 
from models with cosmic values; the energy resolution of the \rosat\ PSPC 
is too low to be 
sensitive to abundance variations, and both GIS spectra do show 
small indication of enhanced magnesium abundances, but the 
large $\chi^2$ resulting from the strong residuals at 
silicon and sulfur line energies renders it difficult to assess its 
significance (results are given in Table.~2b.).
\subsubsection{Study of the ``B'' Region }
\noindent
We have carried out an identical analysis on the ``B''  region 
which, as mentioned above, designates the part of the 
remnant located at the other ``edge'' of G272.2$-$3.2.
We extracted data from an ellipse of 2$^\prime$41$^{\prime\prime}$ and 
5$^\prime$35$^{\prime\prime}$ minor and major axis, and located at 
 09$^{\rm h}$07$^{\rm m}$14$^{\rm s}\!.$5, 
-52$^\circ$06$^\prime$19$^{\prime\prime}\!.$6 (J2000) 
(see Fig~.3 -- left panel).\\
This region selection allows us to study most of the emission within the 
remnant, but encompasses more than one CCD on the SIS detectors. The data
from these detectors had to be combined prior to any analysis.
As in the previous analysis, we used 
data from the \rosat\ PSPC to constrain the value for the column density.
After background subtraction, the count rates are 
$0.19\pm0.003$ cts~s$^{\rm -1}$ for the \rosat\ PSPC and 
$0.075\pm0.001$ cts~s$^{\rm -1}$, 
and $0.066\pm0.001$~cts~s$^{\rm -1}$ in \asca\ SIS and GIS 
respectively.\\
We fit the data sets using an NEI model. Only the normalization varies from
data set to data set. We applied a gain shift equal to that found 
in the previous analysis.
We find ($\chi^2$/$\nu$= 635/317) 
 a column density of $N_{\rm H} = 
 1.30^{\rm +0.03}_{\rm -0.04}\times 10^{\rm 22}$ atoms~cm$^{\rm -2}$, 
 associated with a temperature of $kT = 
0.65^{\rm +0.06}_{\rm -0.02}$ keV and an ionization
timescale of  $4180^{\rm + 960}_{\rm -1290}$cm$^{-3}$ years (see Table.~3. for
a summary).\\
From all the previous analysis, it seems possible to get a 
consistent description of the remnant. 
Both regions ``A'' and ``B'' have compatible temperature and ionization
timescale.
In the following section, we will examine 
what picture of the evolutionary state of the remnant these results imply.
\subsection{Radial Temperature Gradient}
\noindent
One of the important goals of this work is to understand 
 the origin of the centrally-peaked X-ray morphology of the remnant.
The temperature profile is an 
important diagnostic tool to separate between 
models which can explain this kind of morphology. Some models, like a 
one-dimensional, spherically symmetric, hydrodynamic shock code 
\citep{hug84}, 
predict measurable variations of the observed temperature across the 
 remnant, while others do not (White \& Long 1991;  hereafter WL). 
We have  studied temperature variations across the remnant using 
data from the GIS~2. 
 The remnant was separated in 5 annuli centered at 
 09$^{\rm h}$06$^{\rm m}$46$^{\rm s}$, 
-52$^\circ$06$^\prime$36$^{\prime\prime}\!.$14 (J2000) chosen so as to 
contain the same number of events (around 3220 
events) per annulus. 
 We fixed the column density and the ionization timescale 
to the values found in the  previous spectral analysis. 
All elemental abundances are kept linked to each other at their 
nominal ratio. This is only an approximation used to get an 
estimate of the possible temperature variation across the SNR.
In the following section, we will examine both the 
temperature and surface brightness  profile and consider two scenarios 
( evaporation 
of cold clouds in the remnant interior and late-phase evolution incorporating
the effects of thermal conduction) which 
have been applied to other remnants successfully to reproduce both 
the morphology and the temperature profile. 

\subsection{Radial profile of G272.2$-$3.2 }
\subsubsection{Sedov-Taylor solution}
In the soft X-ray band G272.2$-$3.2 is almost perfectly circular 
in appearance with a radius of $8^\prime$. 
As mentioned in the \S 2, the distance to G272.2$-$3.2 is not well known and its 
measured column density is large. In addition, there is no trace of any 
high-energy contribution from a central object.
In this context, we have examined the possibility that G272.2$-$3.2 
is a standard shell-like remnant 
which appears centrally peaked because of the large absorption along 
the line of 
sight or because of projection effects. If there were a density enhancement 
in the shell near the  projected center of the remnant (due for example to 
the possible presence of metal-rich ejecta) that was
similar to the density enhancement at the location of the shell
toward the west, the remnant would have a centrally peaked morphology similar 
to the one observed. The fact that G272.2$-$3.2 presents both 
some limb brightening and centrally peaked emission could support this 
simple explanation. To quantify this model a little bit more 
we have studied the profile of G272.2$-$3.2 in the \rosat\ HRI in 
4 different quadrants of the remnant.
We chose the quadrants so that 
the brightest part of the remnant (that we called 
region ``A'' in our spectral analysis ) belongs to one quadrant only. 
The profiles are shown in Fig.~7 and reveals the distinct enhancement on the 
west-side of the remnant.  We have computed then the expected X-ray emission
in the center for a simple shell model in which the inner radius is taken 
at 5~$\myarcmin$ (instead of the 7.4~$\myarcmin$  expected in the standard 
Sedov-Taylor solutions) to accommodate the measured west enhancement.
We find that the central X-ray emission enhancement 
in the northern quadrant is a factor of 3 to 4 times brighter
than expected from the shell at that position. This requires a 
density about a factor of two higher than in the shell, a factor not excluded
by our analysis. 
In this case, the remnant evolution can simply be described 
by a set of Sedov-Taylor self-similar solutions. \\
The total X-ray emitting volume is
 $V = 1.94\times 10^{59}\,f\,D_{\rm 5}^{\rm 3}$ $\theta_{\rm 8}^{\rm 3}$ 
cm$^{\rm 3}$,   where $f$ is the volume filling
factor of the emitting gas within the SNR, 
$D_{\rm 5}$ is the distance to the remnant in units of 5~kpc, and
 $\theta_{\rm 8}$ the angular radius in units of 8$\myarcmin$. 
In the following discussion, we have used the results of the NEI fit to the
complete remnant (\asca\ GIS and \rosat-PSPC combined; see Table.~1.). 
Because of the relatively large value of the $\chi^2$ found in our best fit 
analysis, we have estimated physical parameters using a larger 
range of values 
than that found by the standard $\chi^2$ analysis 
and given in Table.~1 (we increased the errors by a factor 3). 
For an NEI normalization ranging from 
3.2 to 4.1 $\times$~10$^{\rm 12}$ cm$^{\rm -5}$ and 
a ratio $n_e/n_{\rm H}$ between 1.039 and 1.084,  
we get a hydrogen number  density
 $n_{\rm H}$ between 0.21 and 0.25
 $\,D_{\rm 5}^{-1/2}\,\theta_{\rm 8}^{-3/2}\,f^{-1/2}$ cm$^{-3}$. 
The mass of  X-ray emitting plasma $M_{X}$, 
in a Sedov-Taylor model, is between 35 and 
184$\,D_{\rm 5}^{5/2}\,f^{1/2}\theta_{\rm 8}^{3/2}$ $M_\odot$.
The estimated age of the remnant varies between 6250 and 15250 years
and the initial supernova explosion ranges between 
1.3 and 4.9~$\times 10^{\rm 50}\,D_{\rm 5}^{\rm 5/2}\,\theta_{\rm 8}^{\rm 3/2}\,f^{\rm -1/2}$~ergs.
At a distance of 2~kpc (our lower limit on the distance), 
the emitting X-ray mass value is too small to allow for the remnant to have
reached its Sedov-Taylor phase and the energy explosion is
quite atypical of a supernova event.\\
\noindent 
As indicated in all the quantities derived, all these 
estimates have a strong dependence on the distance to the remnant.
If the remnant is even  further away than the distance derived by 
\citet{gre94}, 
and this may well be the case, considering 
that the distance estimate used there is probably inaccurate 
by at least a factor 2, it is not impossible 
to reconcile the values deduced for both the emitting X-ray mass and the 
initial explosion energy with acceptable estimates for a standard SNR.

\subsubsection{Cloudy ISM}
\noindent
Although it is possible that the centrally peaked X-ray morphology of 
G272.2$-$3.2 is due to the viewing effects mentioned in the previous section, 
we have examined other scenarios which could also explain it. 
In particular we have used a model \citep{whi91} 
based on cloud evaporation. 
This model 
invokes a multi-phase interstellar medium consisting of cool
dense clouds embedded in a tenuous intercloud medium.  The blast wave
from the SN explosion propagates rapidly through the intercloud medium,
engulfing the clouds in the process. In the model, these clouds are
destroyed by gradually evaporating on a timescale set by the saturated
conduction heating rate from the hot post-shock gas. Since this
timescale can be long, it may be possible for cold clouds to survive until
they are well behind the blast wave which, as they evaporate, can
significantly enhance the X-ray emission from near the center of 
the remnant.\\
The timescale for cloud evaporation is one of the two extra 
parameters in the WL model added to the three of 
the standard Sedov solution: explosion energy $E_0$, ISM density
$n$, and SNR age $t$.  This timescale, which is expressed as a
ratio of the evaporation timescale to the SNR age, 
$\tau_e \equiv t_{\rm evap}/t$, nominally depends on different factors, 
such as the composition of
the clumps and the temperature behind the shock front, although such 
dependencies are not included explicitly in the model.  The other extra 
parameter, $C$, represents the ratio of the mass in clouds to the mass
in intercloud material.  For appropriate choices of these 
parameters, the model can reproduce a centrally peaked X-ray emission
morphology - see for example the application of this model to the 
centrally-peaked remnants W28 and 3C400.2 \citep{lon91}.
In the evaporating model atoms from the cold clouds enter the hot medium 
on timescales smaller than the ionization timescale. The line emission occurs
after the ion has left the cloud and the ionization occurs in the hot phase.
In this case, one would expect the highest ionized material to be near 
the center of the remnant or equivalently, to have the 
smallest range of ionization timescales further out. 
In all the following analysis, we have used emissivities derived from 
the NEI model with the ionization timescale fixed to the value 
from the complete remnant analysis. \\
\noindent
We explored the parameter space of the two extra parameters and compare in 
Fig.~8 the results found for 3 representative sets of parameters with the 
HRI profile deduced  for both  the 
northern and western quadrants (as defined in Fig.~7). 
We also compared the expected 
temperature profile for the best value of $C$ and $\tau$  with the \asca\ 
GIS measured temperature profile using an NEI model as described in 
Section 3.3.  
One can see that we can reproduce the
emission weighted temperature variation across the remnant.
The predicted profile agrees within the error bars with the emission profile 
from the most ``centrally peaked'' quadrant of the remnant but disagree with 
the profile extracted from the Western part of G272.2$-$3.2 (shown in dotted lines 
on Fig.~8).
For comparison we have 
kept two other good trials for ($C$, $\tau$) equal to (50,20) 
and (35,10) respectively. 
The WL model allows for variations from the Sedov-Taylor 
in the physical quantities estimated from the spectral analysis but the 
differences stay small. 
For the values of ($C$=100,$\tau$=33) 
which provide an acceptable fit to the
data, $M_{X}$ does not vary by more than 10\% from the value derived in the 
Sedov-Taylor model. 
\subsubsection{Thermal conduction model}
\noindent
Thermal conduction is an alternative model to explain the morphology
and emission from thermal composite remnants, and was used 
most notably for the SNR W44 [see \citet{cox99} for a detailed description 
of the model, and \citet{she99} for its application to W44]. 
It can be summarized as the first attempt to include classical thermal 
conduction in a fully-ionized plasma \citep{spi56}, moderated by 
saturation effects \citep{cow77}, to smooth the large temperature
variations that are created in a Sedov-Taylor explosion.  The result
is a very flat temperature profile, dropping rapidly at the edge of
the remnant. \\ 
\noindent
We used {\sc odin}, a one-dimensional hydrocode which includes the
effects of non-equilibrium cooling and thermal conduction \citep{smi00} 
to create models for a grid of explosion energies,
ambient densities, and ages.  SN explosion energies from
$0.3-2\times10^{\rm 51}$\,ergs were considered along with ambient
densities from 0.05-10 cm$^{\rm -3}$, for ages in the range 2000-50,000~years. 
These give a range of remnant radii, X-ray
spectra, and surface brightnesses that can be compared to the observed
values for the remnant.  As shown in Fig.~8, the best-fit
temperature varies between 0.65-0.8 keV, and the surface brightness
from $0.5-2.5\times10^{\rm -3}$\,cts~s$^{\rm -1}$~arcmin$^{\rm -2}$ 
in the \asca\ GIS.
The average surface brightness is $\sim 10^{\rm -3}$\,cts~s$^{\rm -1}$~arcmin$^{\rm -2}$. \\
\noindent
We found that low ($n_0 \lesssim 0.5$\,cm$^{-3}$) ambient densities
were required in order to get a central temperature of 0.7 keV, 
just as was found in the Sedov-Taylor model.
Higher densities led to lower temperatures (assuming a fixed surface
brightness of $10^{\rm -3}$\,cts~s$^{\rm -1}$~arcmin$^{\rm -2}$), 
independent of the 
explosion energy.  All other SNRs where the thermal
composite model has been used successfully showed much higher ambient
densities [4.72 cm$^{\rm -3}$ for W44 \citep{she99} and a best-fit
density between 1-2 cm$^{\rm -3}$ for MSH~11$-$61{\sl A} \citep{sla00}].
This  low inferred density is a direct result of the high measured 
temperature, substantially larger than in those SNRs.  
In addition, the radial extent of the remnant on the sky is 8$^{\prime}$, 
so the physical radius is 11.6$D_{\rm 5}$\,pc.
The thermal conduction models we examined with the
observed surface brightness and temperature require a minimum radius
of 13.5 pc and age of 12,000 yr, suggesting a minimum distance of 5.8~kpc in 
the case of an explosion energy of $3\times10^{\rm 50}$\,ergs. 
A larger explosion energy will increase the radius and therefore the 
distance estimate. \\
\noindent
We compared the observed GIS spectrum against solar-abundance NEI
models from {\sc odin}\ and the Raymond-Smith \citep{ray77}
plasma model.  Using the above-mentioned explosion parameters, the
Sedov-Taylor swept-up mass is $\approx 120$M$_{\odot}$. The spectrum derived 
in this configuration leads to  a reduced
$\chi^2$ of 5.7 (assuming a column density of 
1.3$\times10^{\rm 22}$\,atoms~cm$^{\rm -2}$). In addition, 
while the model matches the He-like
magnesium lines at 1.4 keV, the model silicon emission at 1.8 keV is
less than half the observed value, and the sulfur lines at 2.46 keV
are far too weak.  Using larger explosion energies only increases this
trend in the models.  This suggests that either the SNe that formed
G272.2$-$3.2 had an atypically-small explosion energy, or that a
thermal-conduction SNR model which does not include ejecta is not a
good fit to the observations.

\section{Discussion \& Summary}
\noindent 
G272.2$-$3.2 presents a couple of puzzles that we have addressed in this
paper. 
First, the X-ray data 
point to a youngish remnant (its size -- at the assumed 
distance of 5~kpc, its X-ray temperature, and the potential
overabundance of Mg, Si and S) although its infrared emission resembles 
an older remnant profile. 
This lack of 12~$\mu$m and 25~$\mu$m emission could be due to 
a low dust density possible if the supernova explosion occurred in 
a cavity. Such a scenario would imply a massive progenitor (more than 8 to 10
M$_\odot$) with strong stellar winds, but is hard to combine 
with the model of a cloudy ISM put forward to explain the morphology.  
Note that according to nucleosynthesis models, a progenitor more massive 
than 15M$_\odot$ would 
yield more than 0.046, 0.071, and 0.023 M$_\odot$  of Mg, Si and S
respectively \citep{thi96}, which would require a 
swept-up mass of more than 100M$_\odot$ to lead to Mg and Si abundances less
than twice that of cosmic abundances. \\
The radio morphology of G272.2$-$3.2 
is difficult to characterize simply because of 
its low surface brightness which prevented a complete mapping of the remnant
by the ACTA. \\
The X-ray peaked morphology can be explained by viewing effects or
a cloud evaporation model. 
Both explanations provide reasonable 
estimates for the initial explosion energy and the total mass swept-up
(depending here again on the distance estimate). A model involving thermal 
conduction, successfully applied to W44 \citep{she99} 
requires the remnant to be around 5000 years old and associated with a 
low SN explosion energy. The model fails to 
reproduce the centrally peaked morphology 
of the remnant, consistent with G272.2$-$3.2 being too young for 
thermal conduction to have taken place. \\
\noindent
The spectrum of the entire remnant presents evidence of strong 
lines (in the residuals of the \asca\ GIS) 
which are not easily accounted for in the 
existing models (both CEI and NEI models). This could be a result of a
a degraded response of the detectors, bad determination of the
 gain applied to the signal or the sign of something more fundamental, linked
either to the simple models used or the properties of the remnant itself. 
We note that such strong lines  
are detected at the same energy range 
in at least two other remnants which present similar 
morphological characteristics: W44 \citep{har97} and 
MSH~11$-$61{\sl A} \citep{sla00}. \\
\noindent
Because of the distance uncertainty, it is difficult to derive the
initial parameters of the remnant (time and initial energy of the explosion).
Low explosion energy supernovae do occur (see Mazzali et al. 1994 and 
Turatto et al. 1996 for the study of SN 1991bg)
but this would be hard to reconcile with a massive progenitor
scenario. \\
\noindent
All these estimates are consistent with G272.2$-$3.2 being in the adiabatic phase
of its expansion, but depend crucially upon the distance to the remnant. 
A better determination of this parameter 
would help characterize unambiguously the evolutionary state of G272.2$-$3.2 and 
will help further constrain the general picture of SNR's evolution. 
\par
\vskip 24 pt

\acknowledgments
Our research made use of data obtained from the High Energy
Astrophysics Science Archive Research Center Online Service, provided
by the NASA/Goddard Space Flight Center. 
POS acknowledges support through NASA grants NAG5-2638 and NAG5-3486,
and contract NAS8-39073.

\clearpage

\begin{figure}
\plotone{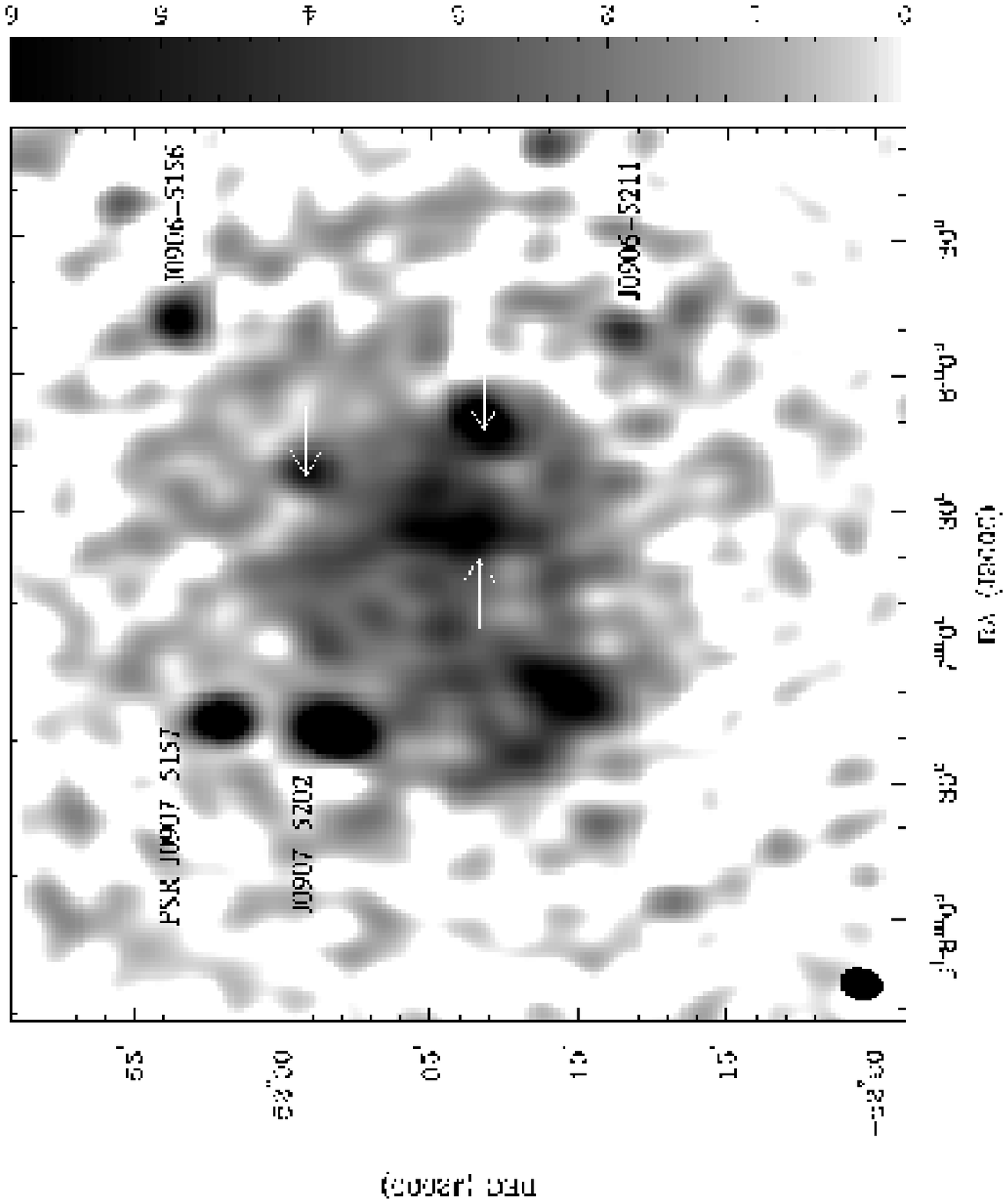}
\caption{Radio image of G272.2$-$3.2 taken with the Australian Telescope Compact 
Array (ATCA) at 1.4~GHz. The optical filaments are indicated with by arrows. 
The positions of nearby pulsars are also indicated (Duncan et al. 1997).\label{fig1}}
\notetoeditor{This figure has to be rotated}
\end{figure}

\clearpage

\begin{figure}
\plotone{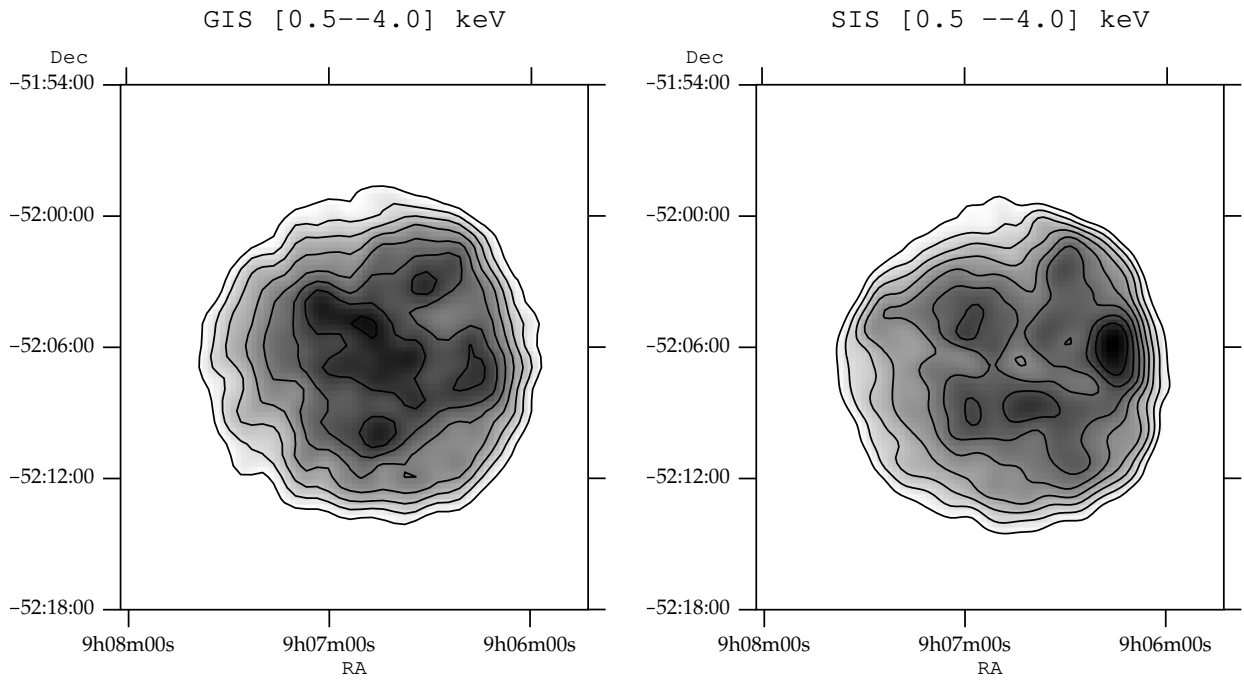}
\caption{\asca\ GIS(left) and SIS(right) images of G272.2$-$3.2 between 0.5 and 4.0~keV.  Contour values  are linearly spaced from 30\% to 90\% of the peak surface brightness in each map. Peak/Background values are (2.47/0.12) for the GIS and (3.20/0.19) for the SIS. All numbers are expressed in units of $10^{-3}$ counts~s$^{-1}$~arcmin$^{-2}$.\label{fig2}}
\end{figure}

\clearpage
\begin{figure}
\plotone{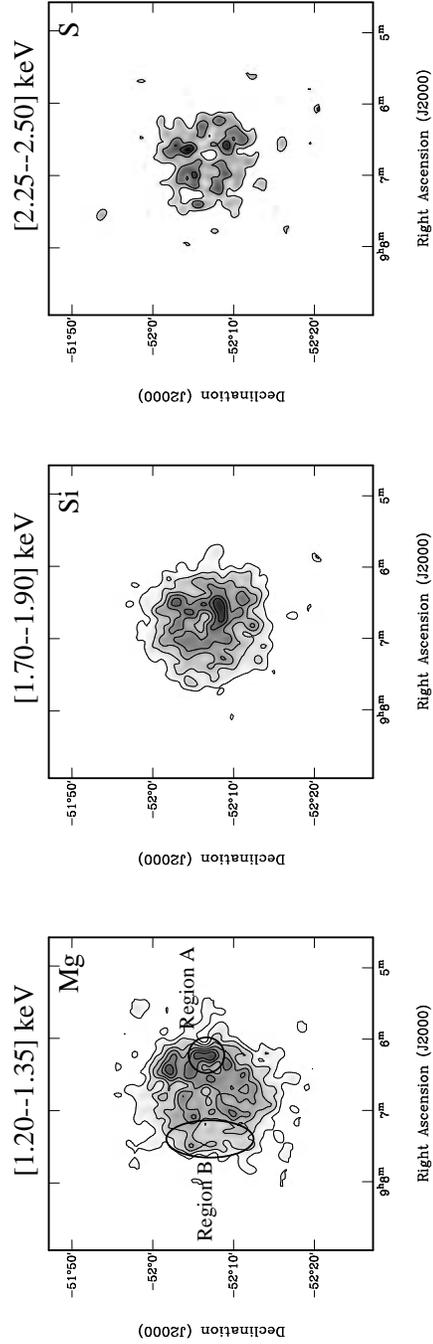}
\caption{ \asca\ SIS images of G272.2$-$3.2 in three narrow band of energy.  The energy band were chosen to isolate emission from Mg XI, Si XIII and S XV. We have indicated by a circle  and an ellipse what is referred to as ``region A'' and ``region B'' in the analysis.\label{fig3}}
\notetoeditor{This figure has to be rotated and reduced in size}
\end{figure}
 
\clearpage
\begin{figure}
\plotone{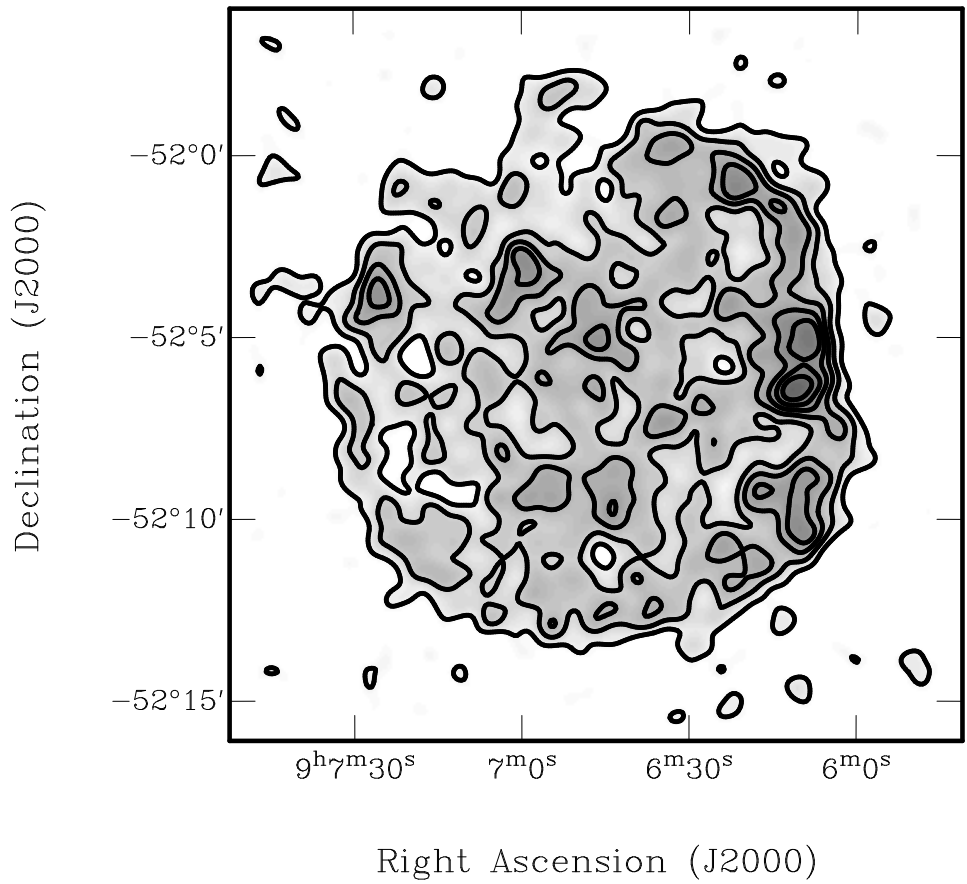}
\caption{\rosat\ HRI image  of G272.2$-$3.2 after standard processing (Snowden et al. 1994).\label{fig4}}
\end{figure}

\clearpage
\begin{figure}
\plotone{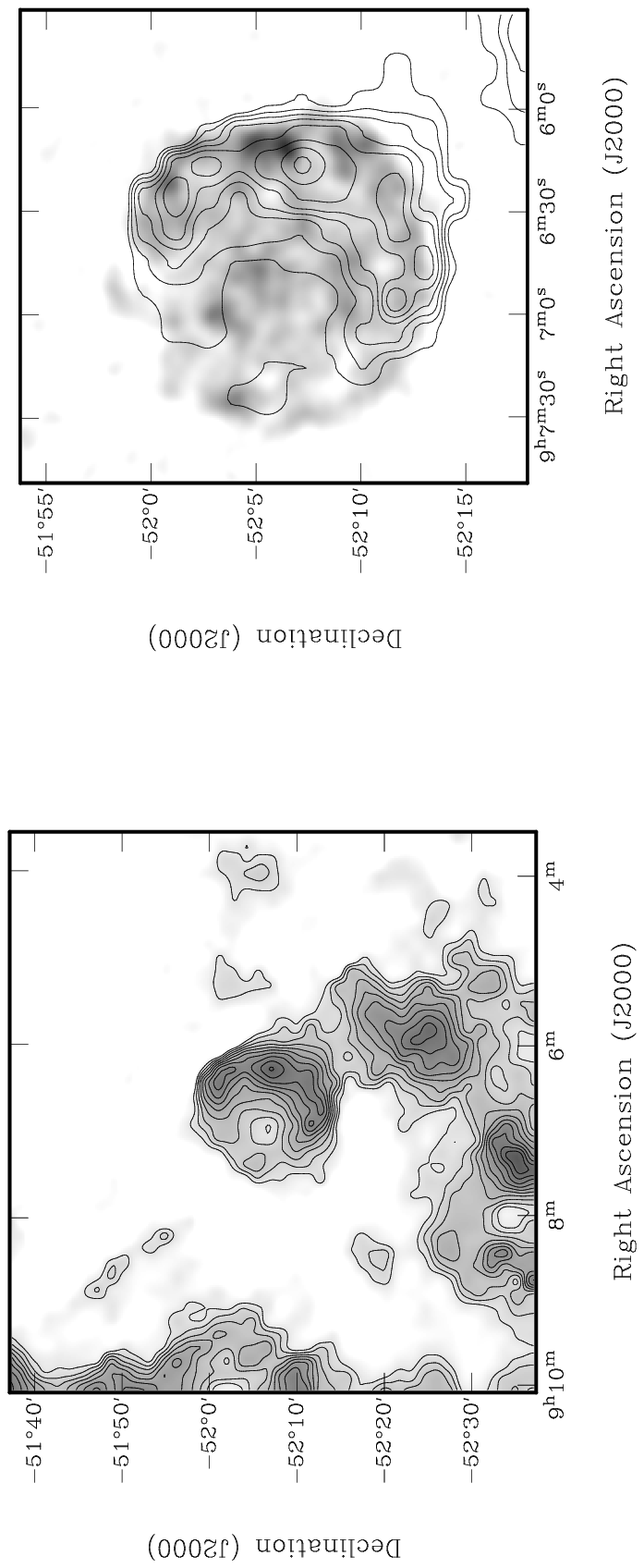}
\caption{{\it IRAS} image of the 60$\mu$m/100$\mu$m emission for G272.2$-$3.2. (right) \rosat\ HRI image superposed with contours from a close-up of the previous image.\label{fig5}}
\notetoeditor{This figure has to be rotated and reduced in size}
\end{figure}

\clearpage
\begin{figure}
\plotone{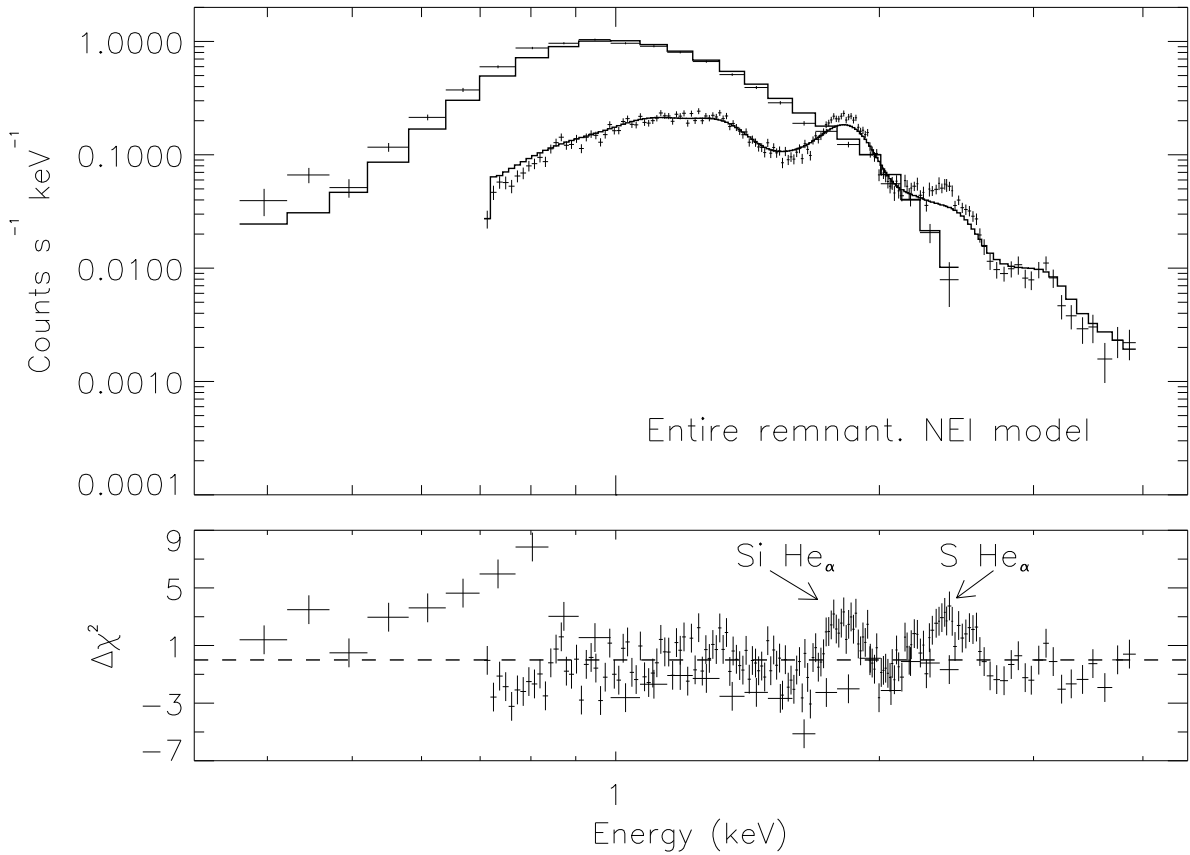}
\caption{Spectra of  the complete remnant for \rosat\ PSPC and \asca\ GIS. 
The two GIS detectors have been merged for display purposes only. 
The solid curves in the top panels show the best-fit NEI thermal plasma; the bottom panels plot the data/model residuals.\label{fig6}} 
\end{figure}

\clearpage
\begin{figure}
\plotone{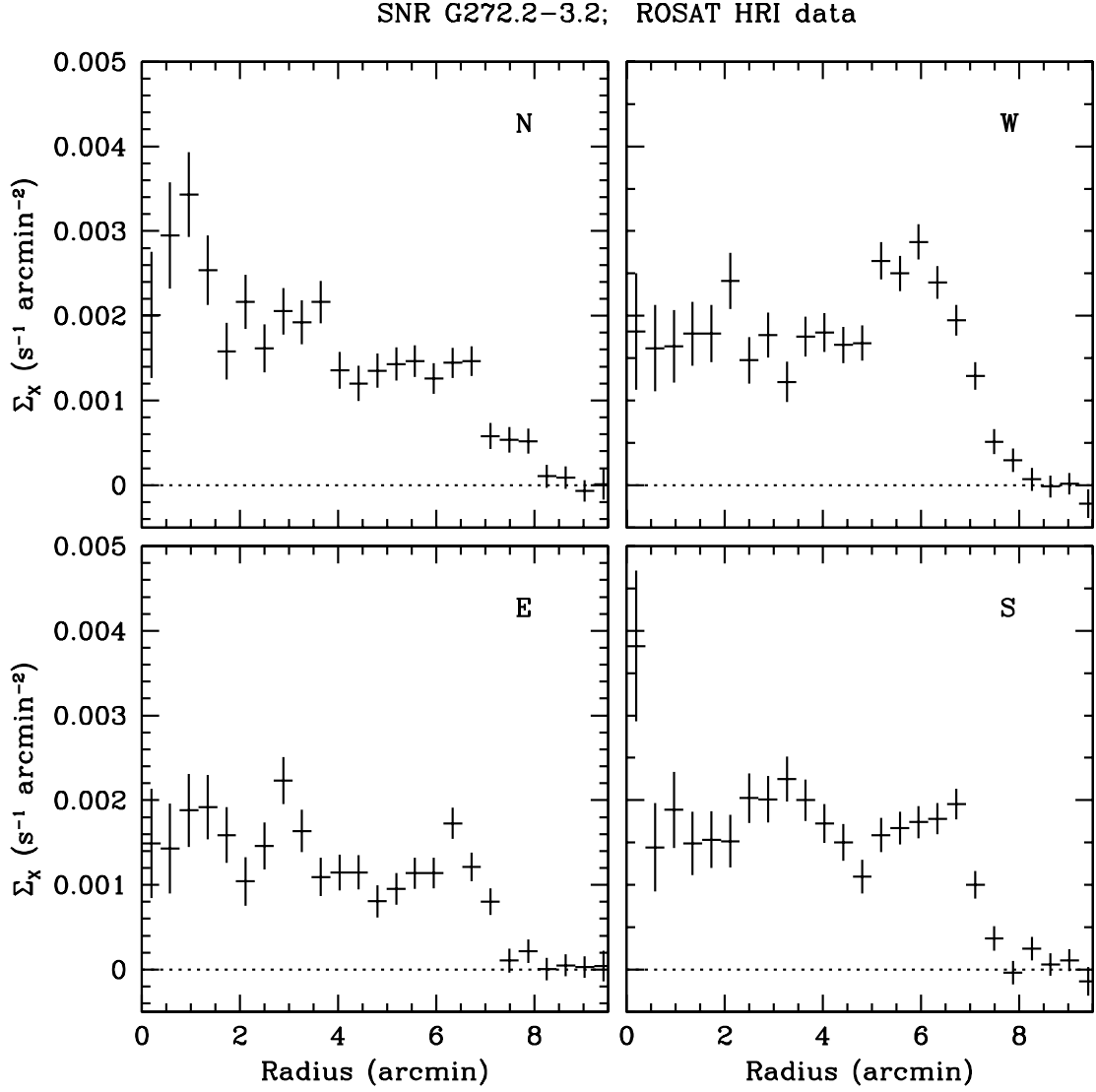}
\caption{\rosat\ HRI profile of the remnant extracted in 4 quadrants - We have chosen the cut so that region A , the region of bright optical filaments,  fall into only one quadrant.\label{fig7}} 
\end{figure}

\clearpage
\begin{figure}
\plotone{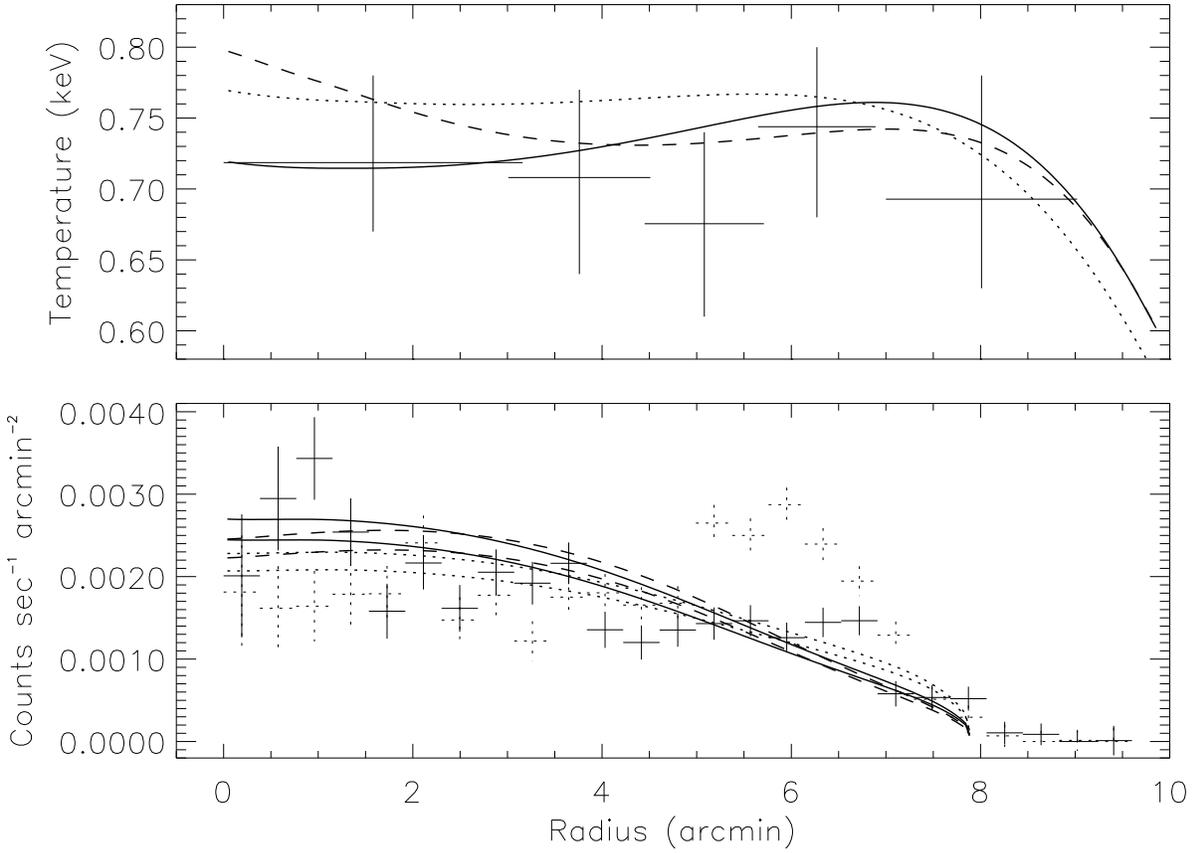}
\caption{Radial temperature variation across the remnant (only up to the first five bins, representing the extent of the remnant) and the associated variation of the surface brightness. The temperature is measured  using the \asca\ GIS . The emission weighted  temperature variation associated with the models is shown as well.  The simulated profile is generated using \rosat\ HRI.  We have shown two extremes results: one obtained for the Northern quadrant (as defined in Fig.~7) shown in solid lines and one for the Western quadrant shown in dotted lines.   We have shown the results from White \& Long simulation using 3 different values of ($C$,$\tau$). Results for ($C$,$\tau$)=(100,33)  are shown in solid lines. Dotted lines are for ($C$,$\tau$)=(50,20) and dashed lines are for ($C$,$\tau$)=(35,10). We have also indicated  the range of normalization allowed by the error bars in the fit. \label{fig8}} 
\end{figure}

\clearpage
\footnotesize
\centerline{\bf Table 1.}
\centerline{\bf Results from the Spectral Analysis}
\vspace{0.5cm}
\centerline{\begin{tabular}{lc} \tableline\tableline \\[-8mm]
\multicolumn{1}{l}{Parameter} &
\multicolumn{1}{c}{Fit results} \\[1.5mm] \tableline 
\multicolumn{2}{c}{Complete remnant (GIS and \rosat\ PSPC), NEI thermal model$^{\rm a}$}\\[1.5mm] \tableline 
$N_{\rm H}$ (atoms~cm$^{-2}$)& 1.12$\pm$0.02$\times$10$^{\rm 22}$\\[1.5mm]
$kT$ (keV) &   0.73$^{\rm +0.03}_{\rm -0.04}$ \\[1.5mm]
log($n_et$) (cm$^{\rm -3}$~s) &   10.83$^{\rm +0.06}_{\rm -0.08}$ \\[1.5mm]
Normalization(cm$^{\rm -5}$)$^{\rm b}$ & (3.5--3.8)$\times$10$^{\rm 12}$    \\[1.5mm] 
Flux (ergs~cm$^{\rm -2}$~s$^{\rm -1}$)~{\rm ([ 0.2 -- 3.0] keV)}  & (2.2--2.5)$\times 10^{-10}$  \\[1.5mm] 
Flux (ergs~cm$^{\rm -2}$~s$^{\rm -1}$)~{\rm ([ 3.0 -- 10.0] keV)}  & (4.3--4.6)$\times 10^{-13}$  \\[1.5mm] 
Flux (ergs~cm$^{\rm -2}$~s$^{\rm -1}$)~{\rm ([ 0.4 -- 2.4] keV)}  & (2.0--2.3)$\times 10^{-10}$  \\[1.5mm] 
$\chi^{2}$/$\nu$ & 691.95/354 \\[1.5mm]\tableline
\multicolumn{2}{l} {$^{\rm a}$ Single-parameter 1~$\sigma$ errors}\\
\multicolumn{2}{l} {$^{\rm b}$ N=(${n_{\rm H}n_eV \over 4\pi D^2}$)}\\
\end{tabular}}

\clearpage
\notetoeditor{Table 2a and 2b are side by side}
\vspace{2in}
\centerline{\begin{tabular}{lc} \tableline\tableline \\[-8mm]
\multicolumn{2}{c}{ Table 2a. Region A, NEI thermal model}\\[1.5mm] \tableline 
\multicolumn{1}{l}{Parameter} &
\multicolumn{1}{c}{Fit results} \\[1.5mm] \tableline 
$N_{\rm H}$ (atoms~cm$^{-2}$)& 1.17$^{\rm +0.02}_{\rm -0.06}\times$10$^{\rm 22}$\\[1.5mm]
$kT$ (keV) &   0.86$^{\rm +0.08}_{\rm -0.22}$ \\[1.5mm]
log($n_et$) (cm$^{\rm -3}$~s) &   10.38$^{\rm +0.32}_{\rm -0.04}$ \\[1.5mm]
Normalization (cm$^{\rm -5}$) & (1.35--3.50)$\times$10$^{\rm 11}$ \\[1.5mm] 
Flux (ergs~cm$^{\rm -2}$~s$^{\rm -1}$)~{\rm ([ 0.2 -- 3.0] keV)} 
& (1.6--4.1)$\times 10^{-11}$  \\[1.5mm] 
Flux (ergs~cm$^{\rm -2}$~s$^{\rm -1}$)~{\rm ([ 3.0 -- 10.0] keV)}
 & (2.3--5.9)$\times 10^{-14}$  \\[1.5mm] 
Flux (ergs~cm$^{\rm -2}$~s$^{\rm -1}$)~{\rm ([ 0.4 -- 2.4] keV)} 
& (1.4--3.6)$\times 10^{-11}$  \\[1.5mm] 
$\chi^{2}$/$\nu$ & 248.35/133 \\[1.5mm] \tableline
\end{tabular}}
\vspace{1in}
\centerline{\begin{tabular}{lc} \tableline\tableline \\[-8mm]
\multicolumn{2}{c}{ Table 2b. Region A, NEI thermal model; Magnesium
Abundance varies freely}\\[1.5mm] \tableline 
\multicolumn{1}{l}{Parameter} &
\multicolumn{1}{c}{Fit results} \\[1.5mm] \tableline 
$N_{\rm H}$ (atoms~cm$^{-2}$)& 0.95$^{\rm +0.07}_{\rm -0.10}\times$10$^{\rm 22}$\\[1.5mm]
$kT$ (keV) &   1.00$^{\rm +0.35}_{\rm -0.08}$ \\[1.5mm]
log($n_et$) (cm$^{\rm -3}$~s) &   10.64$^{\rm +0.37}_{\rm -0.22}$ \\[1.5mm]
Normalization (cm$^{\rm -5}$) & (0.6--1.9)$\times$10$^{\rm 11}$ \\[1.5mm] 
[Mg]/[Mg]$_\odot$ (\rosat; SIS 0\&1; GIS 2\&3) &  $0.5^{\rm +0.8}_{\rm -0.4}$ ;$2.2^{\rm +2.8}_{\rm -0.9}$; $2.7^{\rm +3.5}_{\rm -1.3}$; $1.4^{\rm +1.7}_{\rm -0.7}$; $1.2^{\rm +1.6}_{\rm -0.5}$   \\[1.5mm] 
Flux (ergs~cm$^{\rm -2}$~s$^{\rm -1}$)~{\rm ([ 0.2 -- 3.0] keV)} 
& (0.5--1.5)$\times 10^{-11}$  \\[1.5mm] 
Flux (ergs~cm$^{\rm -2}$~s$^{\rm -1}$)~{\rm ([ 3.0 -- 10.0] keV)}
 & (2.3--7.6)$\times 10^{-14}$  \\[1.5mm] 
Flux (ergs~cm$^{\rm -2}$~s$^{\rm -1}$)~{\rm ([ 0.4 -- 2.4] keV)} 
& (0.46--1.43)$\times 10^{-11}$  \\[1.5mm] 
$\chi^{2}$/$\nu$ & 176.1/129  \\[1.5mm] \tableline
\end{tabular}}
\clearpage

\vspace{1in}
\centerline{\begin{tabular}{lc} \tableline\tableline \\[-8mm]
\multicolumn{2}{c}{Table 3. Region B, NEI thermal model}\\[1.5mm] \tableline 
\multicolumn{1}{l}{Parameter} &
\multicolumn{1}{c}{Fit results} \\[1.5mm] \tableline 
$N_{\rm H}$ (atoms~cm$^{-2}$)& 1.30$^{\rm +0.03}_{\rm -0.04}\times$10$^{\rm 22}$\\[1.5mm]
$kT$ (keV) &   $0.65^{\rm +0.06}_{\rm -0.02}$ \\[1.5mm]
log($n_et$) (cm$^{\rm -3}$~s) &   11.11$^{\rm +0.10}_{\rm -0.15}$ \\[1.5mm]
Normalization (cm$^{\rm -5}$) & (0.7--1.6)$\times$10$^{\rm 12}$ \\[1.5mm] 
Flux (ergs~cm$^{\rm -2}$~s$^{\rm -1}$)~{\rm ([ 0.2 -- 3.0] keV)} 
& (2.9--7.2)$\times 10^{-11}$  \\[1.5mm] 
Flux (ergs~cm$^{\rm -2}$~s$^{\rm -1}$)~{\rm ([ 3.0 -- 10.0] keV)} & (0.7--1.6)$\times 10^{-13}$  \\[1.5mm] 
Flux (ergs~cm$^{\rm -2}$~s$^{\rm -1}$)~{\rm ([ 0.4 -- 2.4] keV)} 
& (2.8--6.7)$\times 10^{-11}$  \\[1.5mm] 
$\chi^{2}$/$\nu$ & 635/317 \\[1.5mm] \tableline
\end{tabular}}

\end{document}